\pdfoutput=1
\makeatletter
\def\input@path{{revtex/}} 
\makeatother
\documentclass[aps,prl,twocolumn,amsmath,amssymb,superscriptaddress,preprintnumbers]{revtex4-2}
\makeatletter
\def\NAT@sort{\z@} 
\makeatother
\usepackage[utf8]{inputenc}
\usepackage[T1]{fontenc}
\usepackage{lmodern}
\usepackage{amsmath,amsfonts,amsthm,amssymb}
\usepackage{graphicx,xcolor}
\graphicspath{{Figures/}} 
\usepackage{microtype}
\usepackage{hyperref}
\hypersetup{hidelinks}
\colorlet{BLF}{red!60!green}
\colorlet{PLF}{blue!70!red}
\colorlet{GV}{black!50!green}
\colorlet{NOTE}{red!50!blue}

\newcommand \bse {\begin{subequations}}
\newcommand \ese {\end{subequations}}
\newcommand \be {\begin{equation}}
\newcommand \ee {\end{equation}} 
\newcommand \bel {\be \label}
\newcommand \bea {}
\def\bea#1\eea{\begin{align}#1\end{align}}

\newcommand \SF {{\textnormal{SF}}} 

\newcommand \coloneqq {\mathrel{\mathop :}\mathrel{\mkern-1.2mu}=} 
\DeclareMathOperator \Tr {Tr}

\DeclareMathOperator \sgn {sgn}
\newcommand \del {\partial}
\newcommand \Hcal {\mathcal H}   
\newcommand \Lcal {\mathcal L}   
\newcommand \RR {\mathbb R}
\newcommand \Ibf {\mathbf I}
\newcommand \Sbf {\mathbf S}
\newcommand \Siso {\mathbf S^{\textnormal{iso}}}
\newcommand \Sani {\mathbf S^{\textnormal{ani}}}
\newcommand \Kcirc {\mathring{K}}
\newcommand \kcirc {\mathring{k}}

\newcommand \tb {t_{\textnormal{b}}}
\newcommand \ts {t_{\textnormal{s}}}

\begin{document}

\title{Universal scattering laws for quiescent bouncing cosmology}

\preprint{CERN-TH-2020-101}

\author{Bruno \surname{Le Floch}}
\email{bruno@le-floch.fr}
\affiliation{Philippe Meyer Institute, Physics Department, \'Ecole Normale Sup\'erieure, 
  PSL Research University, 
  Paris, France.}
 
\author{Philippe G. \surname{LeFloch}}
\email{contact@philippelefloch.org}
\affiliation{Laboratoire Jacques-Louis Lions \& Centre National de la Recherche Scientifique, 
  Sorbonne Universit\'e,  
  Paris, France.}

\author{Gabriele \surname{Veneziano}}
\email{gabriele.veneziano@cern.ch}
\affiliation{CERN, Theory Department, 
  Geneva
  , Switzerland
\&
Coll\`ege de France,
  Paris, France.}

\date{January 2021 / revised version}


\begin{abstract}
Cosmological bounces occur in many gravity theories. We define singularity scattering maps relating large scale geometries before and after the bounce (assuming no BKL oscillations) and encoding microscopic details of the theory. By classifying all suitably local maps we uncover three universal laws: scaling of Kasner exponents, canonical transformation of matter, directional metric scaling. These are indeed obeyed by Bianchi~I bounces in string theory, loop quantum cosmology and modified matter models; our classification then determines how inhomogeneities and anisotropies traverse bounces, and precisely extracts model-dependent degrees of freedom.
\end{abstract}

\maketitle


\section{Three universal laws}

\paragraph{Toward a unification of bouncing scenarios.}

An important class of proposals to resolve the initial singularity problem in cosmology are bouncing scenarios in which the universe undergoes a contracting phase followed by the expanding phase that includes the present time (see~\cite{BrandenbergerPeter}).
Such scenarios have been constructed through various modified gravity theories~\cite{Cesare-bounce}, matter (often a scalar field) violating the dominant energy condition~\cite{Lehners,Bars,Cai}, or quantum gravity effects in string theory~\cite{GasperiniVeneziano1,GasperiniVeneziano2} and loop quantum cosmology~\cite{Ashtekar}.

\enlargethispage{\baselineskip}

Viable bouncing cosmologies include a phase (e.g.,\@ slow contraction~\cite{Cook:2020oaj,Ijjas:2020dws}) responsible for our homogeneous, isotropic and flat universe, a mechanism responsible for the bounce, and should reproduce unmodified Einstein gravity at much larger scales than the duration of the bounce.
We do not attempt to review here the wide literature on these topics, and
we rely exclusively on this last aspect.
In this Letter, we revisit this old problem by {\sl abstracting away all microscopic details} of the model, and describing spacetime at large time scales as two singular solutions to Einstein's equations joined across the bounce.

\paragraph{Classification of junction conditions.}

Many specific junction conditions have been proposed~\cite{SteinhardtTurok2004,BrandenbergerPeter}, e.g., by analytic continuation~\cite{Xue:2014oea,Belbruno:2018uek}.
We encompass them into the notion of {\bf singularity scattering map} of a microscopic theory, which maps a singular contracting solution of Einstein's equations to the expanding solution resulting from the given theory.
This map encapsulates {\sl all the information needed} to describe, at large scales, arbitrarily inhomogeneous and anisotropic bounces for that theory.

Einstein's equations become ultralocal~\cite{BKL,Damour-et-al,Berger:2002st} near a spacelike singularity.
Provided the microscopic theory respects this ultralocality, its singularity scattering map may not involve spatial derivatives.
Einstein's constraint equations restrict the map further.
Our main contribution is to characterize all possible ultralocal singularity scattering maps, in the presence of a scalar field.
For proofs and an application to plane-symmetric cyclic spacetimes, see our companion paper~\cite{LLV-1} together with  Figure~\ref{fig:illustration} below.
Interestingly, our standpoint distinguishes between universal and model-dependent aspects of junction relations.

\paragraph{\bf First law: scaling of Kasner exponents.}
Our classification uncovers three universal laws obeyed by any ultralocal bounce.
First, Kasner exponents scale as
\bel{law1}
(|g|^{1/2} \Kcirc)_{\text{after}} = - \gamma (|g|^{1/2} \Kcirc)_{\text{before}}
\ee
with a \emph{dissipation constant} $\gamma\in\RR$, which involves the spatial metric~$g$ in synchronous gauge, its volume factor $|g|^{1/2}$, and the traceless part $\Kcirc$~of the extrinsic curvature as a $(1,1)$ tensor.

\paragraph{\bf Second law: canonical transformation.}
Matter, modeled away from the bounce as a minimally coupled massless scalar~$\phi$,
undergoes a canonical transformation:
\bel{law2}
\Phi\colon(\pi_\phi,\phi)_{\text{before}} \mapsto (\pi_\phi,\phi)_{\text{after}}
\text{ preserves } d\pi_\phi\wedge d\phi ,
\ee
as explicited in~\eqref{Phi-is-canonical} given below, where $\pi_\phi$ is the momentum conjugate to~$\phi$.
The \emph{matter map}~$\Phi$ further depends on a scalar invariant $\chi\sim(\Tr\Kcirc^3/(\Tr\Kcirc^2)^{3/2})_{\text{before}}$.

\paragraph{\bf Third law: directional metric scaling.}
The metric after the bounce is obtained by a different nonlinear scaling in each proper direction of~$K$, specifically 
\bel{law3}
g_{\text{after}} = \exp\bigl(\sigma_0 + \sigma_1 K + \sigma_2 K^2\bigr) g_{\text{before}} ,
\ee
where $\sigma_0,\sigma_1,\sigma_2$ are arbitrary for scattering maps~\eqref{Siso} with $\gamma=0$
and are explicited in~\eqref{Sani} in terms of $\Phi,\gamma$ for $\gamma\neq 0$.

\paragraph{Model-dependence.}

The three laws are \emph{universal in the renormalization group sense:} they constrain macroscopic aspects of bounces regardless of their origin from different microscopic corrections to Einstein's equations.
Contrarily to field theory universality classes, which depend on finitely many parameters, ultralocal singularity scattering maps depend on a whole map, namely~$\Phi$.

After introducing the set up, the ultralocality assumption, and our classification of singularity scattering maps, we study the maps associated with specific theories: the pre Big Bang scenario, loop quantum cosmology, and some modified matter models.
Our first-principles calculations in homogeneous (but anisotropic) Bianchi~I universes are consistent with the universal scattering laws \eqref{law1}--\eqref{law3}, which suggests that these theories respect ultralocality.
The ultralocality conjecture near a singularity~\cite{BKL} is known numerically to hold in a slow contraction phase~\cite{Ijjas:2020dws} and we argue in this text that it may hold through the bounce.


\begin{figure}
  \includegraphics{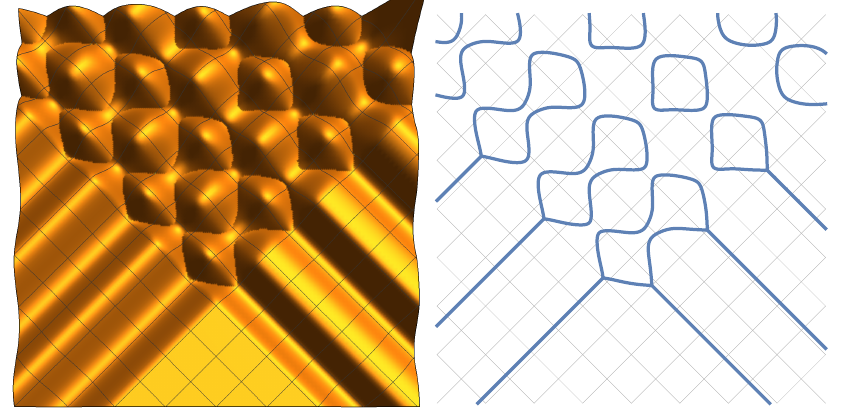}

  \caption{\label{fig:illustration}
    Cyclic spacetime arising from colliding plane gravitational waves, in null coordinates (the future is up)~\cite{LLV-1}.
    Left: area element $A$ of plane-symmetry orbits, depicted as the height of spacetime ``bubbles''.
    Right: singular locus $A=0$ across which we apply the junction $(g^+\!,k^+\!,\phi_0^+,\phi_1^+) = (e^{2(k^--1/3)} g^-\!, \allowbreak k^-\!,\phi_0^-, \phi_1^- + \phi_0^-)$.
    For this example of junction, the evolution problem is well-posed.}
\end{figure}

\section{Singularity scattering maps}

\paragraph{ADM formalism.}

We focus on bouncing scenarios in which corrections to Einstein gravity are negligible away from the bounce locus, which we model as a spacelike singularity hypersurface~$\Hcal$ 
(cf.~\cite{LLV-1} for timelike~$\Hcal$), but are essential at (small) time scales $\tb$ around~$\Hcal$.
Provided spatial inhomogeneities are mild (see below) the spacetime, on each side, is well described at larger time scales by a solution of the Einstein equations which is singular along~$\Hcal$, and these two solutions are connected using a suitable junction.
We model matter as a scalar field that is minimally-coupled and massless away from the bounce.

We work with a Gaussian (or synchronous gauge) foliation in which the metric reads $g^{(4)}=-dt^2+g(t,x)$, the bounce hypersurface being at the proper time $t=0$.
Each constant-time hypersurface is endowed with a Riemannian metric $g=g_{ab}$ and an extrinsic curvature $K=K_a^b$ such that $K_{ac}=K_a^b g_{bc}$ is symmetric. Here, $a,b,\dots$ are local coordinate indices on each time slice.
We consider  the ADM formulation of Einstein's equations:
\bel{ADM}
\aligned
- \del_t^2 \phi + \Tr(K)\, \del_t \phi & = - \, \nabla_a\nabla^a \phi, 
\\
\del_t g_{ab} + 2 \, K_{ab} & = 0,
\\
\del_t K_a^b - (\Tr K)  K_a^b  & =  R_a^b - \del_a\phi \del^b\phi,
\\
(\Tr K)^2 - \Tr (K^2) - (\del_t\phi)^2 & = - R + \del_a\phi\del^a\phi,
\\
\nabla_a K^a_b  - \del_b (\Tr K) + \del_t\phi \del_b\phi & = 0,
\endaligned
\ee
three evolution and two constraint equations.
(We normalize speed of light and Newton constant as $c=8\pi G=1$.)

\paragraph{Quiescent regime.}

The Einstein vacuum equations, \eqref{ADM} with $\phi=0$, are expected to exhibit BKL oscillations as different directions repeatedly expand and contract~\cite{BKL,Damour-et-al,Berger:2002st}.
The matter field~$\phi$ allows all directions to contract hence suppresses oscillations (while allowing anisotropies), leading to quiescent cosmological singularities, a class identified by Barrow~\cite{Barrow}.  Our work focuses on this regime.

Near the singularity, asymptotic profiles describing the main behavior of a solution are found~\cite{AnderssonRendall,Damour-et-al} by neglecting spatial derivatives compared to $t$~derivatives, namely neglecting right-hand sides of~\eqref{ADM}.  This family of \emph{asymptotic profiles} (denoted by a~$*$ subscript) reads
\bel{equa:time-asymptotic-profile}
\aligned
  g_*^{\pm}(t) &= e^{2(\ln|t/t_*|)k^{\pm}}g^{\pm}, 
  \quad & K_*^{\pm}(t) = - (1/t) k^{\pm} , 
  \\
  \phi_*^{\pm}(t) &= \phi_0^{\pm}\ln|t/t_*| + \phi_1^{\pm} , &
\endaligned
\ee
parametrized by \emph{singularity data} $(g^{\pm},k^{\pm},\phi_0^{\pm},\phi_1^{\pm})$ prescribed on each side $\pm=\sgn(t)$ of the bounce.
The asymptotic metric is expressed in terms of the matrix exponential of~$(k^{\pm}_{}{}_a^b)$,
and $t_*>0$ is a time scale.

The singularity data must satisfy the constant trace relation $\Tr k^{\pm}=1$ together with an asymptotic form of the Hamiltonian and momentum constraints
(where $\nabla^{\pm}$ is the connection associated with $g^{\pm}$)
\bel{equa-const-singu}
1 - k^{\pm}_{}{}^a_b  k^{\pm}_{}{}_a^b = (\phi_0^{\pm})^2,
\quad
\nabla^{\pm}_a k^{\pm}_{}{}^a_b
= \phi_0^{\pm} \, \del_b \phi_1^{\pm} .
\ee
In our context, a data set $(g^{\pm}, k^{\pm}, \phi_0^{\pm}, \phi_1^{\pm})$ is called \emph{quiescent} if Kasner exponents $k_i^{\pm}$ (eigenvalues of~$k^{\pm}$) are positive.
The trace and Hamiltonian constraints read $\sum_i k_i^{\pm}=1$ and $1-\sum_i(k_i^{\pm})^2=(\phi_0^{\pm})^2$, respectively.
Asymptotic profiles are generally not exact solutions.

\paragraph{Validity of asymptotic profiles.}

Asymptotic profiles are defined for all times $\pm t\in(0, \infty )$, but are only good approximations in some range $\tb\ll|t|\ll\ts$: indeed,
curvature generically blows up as $t\to 0^{\pm}$ so that
 (e.g.~higher-curvature) corrections become important at a small time $|t|\simeq\tb$,
while the spatial derivatives neglected in~\eqref{ADM} stop being negligible at some large time scale~$\ts$ since they decay slower than time derivatives at $|t| \to  \infty$.

Our assumption of \emph{mild spatial inhomogeneities} is that $\tb\ll\ts$.
Equivalently, we require that at time $\tb$ (bounce duration), spatial derivative terms in~\eqref{ADM} such as $\nabla_a \nabla^a \phi/\phi$, 
$\del_a\phi\del^a\phi$ or~$R$ are parametrically smaller than the typical scale $1/\tb^2$ of the left-hand sides, so that they remain smaller on some time interval $(\tb,\ts)$.
Under this assumption we retrieve the data for the asymptotic profile as the (approximately constant for $\tb\ll|t|\ll\ts$) values
\begin{align}\label{equa-limits}
& (g^\pm, k^\pm, \phi_0^{\pm}, \phi_1^{\pm})
\\\nonumber
& \coloneqq \bigl(|t/t_*|^{2 t K} g,\; -t K,\; t \del_t \phi,\; \phi - t \ln |t/t_*|  \del_t \phi \bigr)_{\tb\ll|t|\ll\ts} .
\end{align}
In the idealized cases $\tb=0$ (singular bounce) or $\ts=\infty$ (spatially homogeneous case) the singularity data can be defined as $t\to 0^{\pm}$ or $t\to\pm\infty$ limits of~\eqref{equa-limits}, respectively.


\paragraph{The new notion.}

We denote by $\Ibf(\Hcal)$ the set of all singularity data $(g^{\pm},k^{\pm},\phi_0^{\pm},\phi_1^{\pm})$ obeying~\eqref{equa-const-singu}.
We define a {\bf singularity scattering map} as a local diffeomorphism-covariant map $\Sbf\colon\Ibf(\Hcal)\to\Ibf(\Hcal)$.
By general covariance and locality, $\Sbf$~is characterized by its effect on any small ball, so the topology of~$\Hcal$ is irrelevant.
We also introduce the corresponding {\bf junction condition}
$(g^+, k^+, \phi_0^+, \phi_1^+) = \Sbf(g^-, k^-, \phi_0^-, \phi_1^-)$.
See Figure~\ref{fig:illustration} for an example map, and an application.

\paragraph{Mathematical advances.}

The existence of solutions to Einstein's equations asymptotic to quiescent profiles~\eqref{equa:time-asymptotic-profile} and satisfying the junction conditions 
is proven in the companion paper~\cite{LLV-1} based on the earlier work~\cite{AnderssonRendall,Damour-et-al}. 
We also refer to~\cite{LeFlochLeFloch-1,LeFlochLeFloch-2,LeFlochLeFloch-3,LeFlochLeFloch-4} for recent progress on the theory of weak solutions with singularities.
Our definition is a generalization to singularity hypersurfaces of Israel's junction conditions~\cite{Israel} for hypersurfaces across which the metric remains regular.
Our junction conditions are reminiscent of \emph{kinetic relations} for phase boundaries in fluid dynamics and material science~\cite{LeFloch-Oslo,LeFlochLeFloch-3}.



\section{A classification of bouncing laws}

\paragraph{Ultralocality.}

As observed in~\cite{BKL,AnderssonRendall,Damour-et-al}, the massless scalar field~$\phi$ suppresses BKL oscillations, so that spatial derivatives can be neglected near a quiescent singularity of Einstein's equations: each spatial point undergoes an (almost) independent evolution.
We assume that the microscopic physics responsible for the bounce does not spoil this decoupling of spatial points.
Namely, we focus on \emph{ultralocal scattering maps,} for which the value of $(g^+,k^+,\phi_0^+,\phi_1^+)$ at a point~$x$ of~$\Hcal$ depends on $(g^-,k^-,\phi_0^-,\phi_1^-)$ at~$x$ but not on (spatial) derivatives.

The ultralocality assumption is supported, for microscopic theories with higher-curvature or higher-derivative corrections, by checking that spacetime invariants are dominated by time derivatives.
For instance, based on the asymptotic profiles~\eqref{equa:time-asymptotic-profile} (henceforth we set $t_*=1$),
\[
|d\phi|_{g^{(4)}}^2 = - (\del_t\phi)^2 + g^{ab} \del_a\phi \del_b\phi \sim - \phi_0^2 |t|^{-2}
\]
where the spatial part is negligible because $g^{ab}\lesssim|t|^{-2k_{\max}}$,
where $k_{\max}$ is the largest Kasner exponent, and generically $k_{\max}<1$ due to~\eqref{equa-const-singu}.
As another example, spatial gradients are also negligible in the Kretschmann scalar
$
|\text{Riem}|_{g^{(4)}}^2\!\!
=\! \tfrac{4}{9}(1 - r^2 - 2 \chi r^3 + 2 r^4)/|t|^4
+ O({\log^4}|t|/|t|^{4k_{\max}})
$
where we defined $\chi^{\pm}=(9/2)\Tr(\kcirc^{\pm}/r^{\pm})^3\in[-1,1]$ and
$r^{\pm}=r(\phi_0^{\pm})= \bigl(1-\tfrac{3}{2} (\phi_0^{\pm})^2\bigr){}^{1/2} = \bigl(\tfrac{3}{2}\!\Tr(\kcirc^{\pm})^2\bigr){}^{1/2}\in[0,1]$.

One may also test ultralocality in a given theory by numerical calculations far from a homogeneous universe, similar to how ultralocality was observed in~\cite{Ijjas:2020dws} in a supersmoothing phase preceeding a possible bounce.

\paragraph{Consequences.}

Ultralocality forces $\kcirc^+=\sum_n\beta_n(\kcirc^-)^n$ for some functions~$\beta_n$ of scalar invariants $\phi_0^-,\phi_1^-,\chi^-$.
As we prove in~\cite{LLV-1}, the constraint $\nabla^+k^+=\phi_0^+\del\phi_1^+$ in~\eqref{equa-const-singu} {\sl can  be fulfilled for all} $(g^-,k^-,\phi_0^-,\phi_1^-)$ {\sl only if}
 $\kcirc^+\sqrt{g^+}=\gamma\kcirc^-\sqrt{g^-}$, our law~\eqref{law1}.  We eventually prove that
\emph{any ultralocal scattering map is either an {\bf isotropic map} $\Siso_{M,\varphi,\epsilon}$ \eqref{Siso} or an {\bf anisotropic map} $\Sani_{\Phi,\gamma}$ \eqref{Sani}}.

\paragraph{Isotropic ultralocal scattering.}

For $\gamma=0$, the constraints~\eqref{equa-const-singu} fix $|\phi_0^+|$ and make $\phi_1^+$ constant, but leave the spatial metric arbitrary:
\bel{Siso}
\aligned
\Siso_{M,\varphi,\epsilon} \colon (g^-,k^-,\phi_0^-,\phi_1^-) \mapsto (g^+,k^+,\phi_0^+,\phi_1^+) 
\\[-.2\baselineskip]
= 
\bigl( \exp(M) g^-, \ \delta/3, \ \epsilon\sqrt{2/3}, \ \varphi \bigr)
\endaligned
\ee
for any constant $\varphi\in\RR$, sign $\epsilon=\pm 1$, and
any linear combination $M=\sum_{n=0}^{2}M_n(\phi_0^-,\phi_1^-,\chi)(k^-)^n$.

The isotropic scattering map $\Siso_{M,\varphi,\epsilon}$ physically describes  an irreversible bouncing scenario in which almost all information is lost:
(i)~Since $k^+=\delta/3$, 
the bounces produce an \emph{isotropic and homogeneous expansion}. 
(ii)~The matter field is constant in space.
(iii)~However, the metric is scaled differently along different eigenvectors of $k^-$.

\paragraph{Anisotropic ultralocal scattering.}
For $\gamma\neq 0$, constraints restrict the matter map~$\Phi$ to be a canonical transformation with $\epsilon=\sgn\gamma$; see~\eqref{Phi-is-canonical} below.
Explicitly,
$\Sani_{\Phi,\gamma} \colon (g^-,k^-,\phi_0^-,\phi_1^-) \mapsto (g^+,k^+,\phi_0^+,\phi_1^+)$
reads
\begin{gather}\label{Sani}
(\phi_0^+, \phi_1^+) = \Phi(\chi,\phi_0^-,\phi_1^-) , \quad
\kcirc^+ = \epsilon (r^+/r^-) \kcirc^- ,
\\[-.3\baselineskip]\nonumber
g^+ = \Bigl|\frac{\gamma r^-}{r^+}\Bigr|^{\frac{2}{3}}
\!\exp\biggl(\! \frac{3\kcirc^-}{2r^-} (\xi-\chi\sigma) + \sigma \Bigl(\frac{9(\kcirc^-)^2}{2(r^-)^2}-\delta\Bigr) \biggr) g^- ,
\end{gather}
where $\del_{\phi_0^-}\xi = - 2 \epsilon (\phi_0^+/r^+)\del_{\phi_0^-}\phi_1^+$ and $\xi$~vanishes at $\phi_0^-=\pm\sqrt{\scriptstyle 2/3}$, and $\sigma = 3(\del_\chi \xi + 2 \epsilon (\phi_0^+/r^+) \del_\chi \phi_1^+)$.

\begin{figure}
  \includegraphics{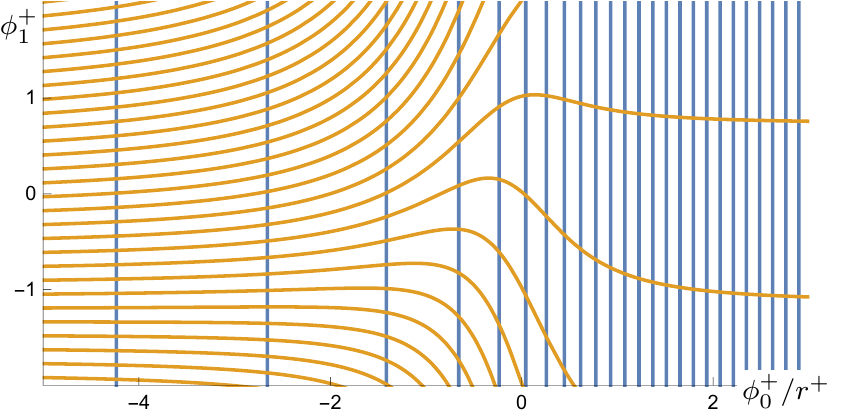}
  
  \caption{\label{fig:pre-Big-Bang}
    Image of equally-spaced constant-$(\phi_0^-/r^-)$ (vertical lines) and constant-$\phi_1^-$ (curved lines) under the matter map~$\Phi$ of the Pre Big Bang scenario~\eqref{pre-Big-Bang-phi} ($\beta^+=-\beta^-$, $u_+=u_-$).
    It preserves $d(\phi_0^{\pm}/r^{\pm})d\phi_1^{\pm}$ so each region here has the same area.}
\end{figure}

\paragraph{The matter map.}

Constraints also imply the law~\eqref{law2}, that
$\Phi$~is a canonical transformation at fixed $\kcirc^-/r^-$.
It preserves
$d\bigl(\phi_0/r(\phi_0)\bigr)\wedge d\phi_1 = d\phi_0\wedge d\phi_1/r(\phi_0)^3$ up to the sign $\epsilon=\sgn\gamma$ (see example in Figure~\ref{fig:pre-Big-Bang}):
\bel{Phi-is-canonical}
\det\biggl(\begin{matrix}
\del_{\phi_0^-}\!(\phi_0^+/r^+) & \del_{\phi_0^-}\phi_1^+ \\
\del_{\phi_1^-}\!(\phi_0^+/r^+) & \del_{\phi_1^-}\phi_1^+
\end{matrix}\biggr)\!
= \epsilon \del_{\phi_0^-}\!\Bigl(\frac{\phi_0^-}{r^-}\!\Bigr)\! = \! \frac{\epsilon}{(r^-)^3} .
\ee 
For $\Siso_{M,\varphi,\epsilon}$, the constant map~$\Phi$ sits at a singular point of the symplectic form~\eqref{Phi-is-canonical} but it is a limit of canonical transformations.
As $\gamma\to 0$ with fixed $\phi_1^+/\gamma$ and $\gamma\phi_0^+/r^+$, $\Sani_{\Phi,\gamma}$ tends towards an isotropic map $\Siso_{M,\varphi,\epsilon}$ with restrictions on the metric factor~$M$.
These maps with $k^+\simeq\delta/3$ are good candidates to approximate supersmoothing models.


\section{Selected examples of bounces}
 
\paragraph{Reduction to Bianchi I.}

We now exhibit the two features~\eqref{law1} and~\eqref{law2} for singularity scattering maps of several models (pre Big Bang, modified matter, etc.\@) in spatially homogeneous bounces.
As we have argued, these laws and~\eqref{law3} can also be derived model-independently from an ultralocality assumption, without spatial homogeneity.
Now, though, we work with a ($d=3$) Bianchi~I metric
\bel{BianchiI}
g^{(4)} = -dt^2 + \omega(t)^{2/3} {\textstyle\sum_{i=1}^3} e^{2\alpha_i(t)} dx^i dx^i,
\ee
with anisotropic stress parameters $\alpha_i$ summing to zero and volume factor $\omega\coloneqq|g|^{1/2}$.

\paragraph{Asymptotic profiles.}

As explained before~\eqref{equa-limits}, spatial homogeneity means that $\ts=  \infty$ namely the bounce is well-described for all $|t|\gg\tb$ by the asymptotic profiles~\eqref{equa:time-asymptotic-profile}, which are exact Bianchi~I solutions to Einstein's equations with a free scalar field.
Explicitly, in the notation~\eqref{BianchiI} we consider bounces that are asymptotic to
\begin{align}\label{Einstein-sols}
\omega & = \pm\omega_0^{\pm}(t-t_0^{\pm}) , \quad
\phi = \phi_0^{\pm} \ln\lvert t-t_0^{\pm}\rvert + \phi_1^{\pm} ,\! \\\nonumber
\alpha_i & = (k_i^{\pm}-\tfrac{1}{3}) \ln\lvert t-t_0^{\pm}\rvert + \nu_i^{\pm} , \quad
(\phi_0^{\pm})^2 + |k^{\pm}|^2 = 1
\end{align}
at $t\to\pm  \infty$,
for some constants $(t_0^{\pm},\omega_0^{\pm},k_i^{\pm},\nu_i^{\pm},\phi_0^{\pm},\phi_1^{\pm})$ such that $\sum_i\nu_i^{\pm}=0$, the Kasner exponents $k_i^{\pm}$ (eigenvalues of~$k^{\pm}$) sum to~$1$, and $|k^{\pm}|^2=\Tr(k^{\pm})^2$.

We are interested in the map that relates parameters describing the two limits.
Invariance under time translations and coordinate redefinitions of each~$x^i$ ensures that
$t_0^-,\ln\omega_0^-,\nu_i^-$ appear precisely as shifts of $t_0^+,\ln\omega_0^+,\nu_i^+$, respectively, so $(t_0^+-t_0^-,\omega_0^+/\omega_0^-,\nu_i^+-\nu_i^-,k_i^+,\phi_0^+,\phi_1^+)$ only depend on $(k_i^-,\phi_0^-,\phi_1^-)$.
For brevity we focus here on $(\omega_0^+/\omega_0^-,k_i^+,\phi_0^+,\phi_1^+)$ and not the metric and time offset.

The first scattering law~\eqref{law1} that we will verify in concrete bounces translates, in Bianchi~I notations, to
$\omega_0^+(k_i^+-\tfrac{1}{3})=\gamma\omega_0^-(k_i^--\tfrac{1}{3})$ for some $\gamma\in\RR$.
In particular, $\omega_0^+ r(\phi_0^+)=|\gamma|\omega_0^- r(\phi_0^-)$.
The second law states~\eqref{Phi-is-canonical}.


\paragraph{Pre Big Bang scenario.}

Our first concrete model is a singular bounce inspired from string theory~\cite{VenezianoSFD,GasperiniVeneziano1,GasperiniVeneziano2},
described by string frame fields $\phi_\SF$, $g_\SF^{(4)}$ obeying suitably truncated metric-dilaton equations.
Bianchi~I solutions related by scale-factor duality are glued along $t_\SF=0$,
assuming that higher derivative and/or higher loop corrections resolve the singularity:
they are $\phi_\SF = \ln|g_\SF^{(4)}|^{1/2} - \ln|t_\SF|$ and
$g_\SF^{(4)} = -dt_\SF^2 + {\textstyle\sum_{i=1}^3} e^{2u_{i\pm}} |t_\SF|^{2\beta_{i\pm}} dx^i dx^i$
on both sides $\pm t_\SF>0$.
The constants $u_{i\pm},\beta_{i\pm}$ obey $\sum_i\beta_{i\pm}^2=1$ and each $\beta_{i+}^2=\beta_{i-}^2$.
Only $\beta_{\pm}$ and differences $u_{i+}-u_{i-}$ are coordinate-invariant;
they depend on~$\beta_-$ and how the singularity is resolved.

The Einstein frame metric $g^{(4)} = e^{-\phi_\SF} g_\SF^{(4)}$\!, proper time~$t$, and canonically normalized scalar $\phi=\phi_\SF/\sqrt{2}$ then take the form~\eqref{Einstein-sols} with, in particular, $\omega_0^\pm(k_i^\pm - {1 \over 3}) = \beta_{i\pm} - {1 \over 3} \sum_j \beta_{j\pm}$. 
Among the $2^3$ choices of~$\beta_+$ allowed by scale-factor duality,
$\beta_+=-\beta_-$ gives an interesting junction, with $\omega_0^+(k_i^+-\tfrac{1}{3})=-\omega_0^-(k_i^--\tfrac{1}{3})$,
\bel{pre-Big-Bang-phi}
\phi_0^+ = - \bigl(2 \sqrt{2} + 4 \phi_0^-\bigr)\!\!\bigm/\!\!\bigl(4 + 3\sqrt{2} \phi_0^-\bigr) ,
\ee
and $\phi_1^+=(r^+/r^-)\phi_1^-+f(\beta_-)$ where the function~$f$ depends on how the singularity is resolved. 
Interestingly, regardless of~$f$ \emph{both laws \eqref{law1} and~\eqref{law2} are obeyed.} 
The matter map~$\Phi$ is depicted in Figure~\ref{fig:pre-Big-Bang}.

All other sign choices (except the trivial $\beta_+=\beta_-$) violate these laws.
By our general classification above, this means that the corresponding junction conditions would not extend to inhomogeneous spacetimes (specifically, applying the transformation pointwise would violate the momentum constraint).


\paragraph{Modified gravity and loop quantum cosmology.}

Both in loop quantum cosmology~\cite{AshtekarWilsonEwing,Wilson-Ewing-LQC} and in quite general modified gravities~\cite{Cesare-bounce} (Brans--Dicke theory, kinetic gravity braiding, mimetic gravity, etc.), the densitized shear $\Kcirc\sqrt{g}$ is continuous (up to a sign) across Bianchi~I bounces.
To derive this, the authors of~\cite{Cesare-bounce} assumed that modifications of gravity are encapsulated in an effective stress-tensor,
preserve spatial rotation invariance, and are strong enough to lead to a bounce but are negligible away from it.

This is precisely our first universal scattering law~\eqref{law1} (with $\gamma=-1$),
which we have proven \emph{without any symmetry assumption.}
It would be very interesting to determine the precise scattering maps for some models and check our second scattering law~\eqref{law2} directly.

\begin{figure}
  \includegraphics{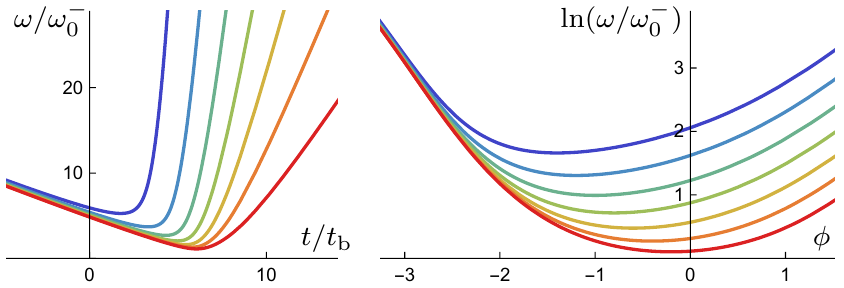}

  \caption{\label{fig:ghost}Bianchi~I symmetric modified matter bounces with Lagrangian $\Lcal=\frac{1}{2}\dot\phi^2-2|\dot\phi|e^{-\phi^2/2}/\tb+e^{-\phi^2}/\tb^2$ for fixed $t_0^-,\phi_0^-$, and $\omega_0^-$ (normalized to~$1$).  Each color corresponds to one value of~$\phi_1^-$, which affects $t\to+\infty$ asymptotics $\omega\simeq\omega_0^+(t-t_0^+)$ and $\phi\simeq\phi_0^+\ln\omega+(\phi_1^+-\phi_0^+\ln\omega_0^+)$ manifest in the two plots.}
\end{figure}

\paragraph{Bounces with modified matter.}

Consider now Einstein gravity coupled to a scalar field with Lagrangian $\Lcal(\phi,X)$ where $X=-|\nabla\phi|^2=\dot\phi^2$.
It is beyond the scope of this Letter to analyse which specific models lead to bouncing solutions (exemplified in Figure~\ref{fig:ghost});
such bounces arise with ekpyrotic matter \cite{KOST}, ghost condensates, Brans--Dicke theory in Einstein frame, etc.
For our setting, Bianchi~I solutions should asymptote to free scalar ones~\eqref{Einstein-sols} at $t\to\pm  \infty$, where $X\to 0$ and $|\phi|\to  \infty$,
so we demand $\Lcal\simeq X/2$ (free scalar) in these limits.

In Bianchi~I spacetimes~\eqref{BianchiI}, the action (per comoving volume) is
$
S = \int\big(\Lcal(\phi,\dot\phi^2) - \dot\omega^2/(3\omega^2) + \frac{1}{2} 
\sum_i \dot\alpha_i^2 \big)\omega\,dt 
$.
As observed in~\cite{Cesare-bounce}, the equation of motion $\del_t(\omega\dot\alpha_i)=0$ for~$\alpha_i$ states that $\lambda_i=\omega\dot\alpha_i$ are constants so their $t\to\pm  \infty$ limits $\pm\omega_0^{\pm}(k_i^{\pm}-\tfrac{1}{3})$ coincide.
This proves in such spacetimes our first scattering law~\eqref{law1} with $\gamma=-1$, for any modified matter Lagrangian~$\Lcal$ that exhibits bounces.

Next, we switch to the Hamiltonian formalism with momenta $\pi_\phi=2\omega\dot\phi\del_X\Lcal$, $\pi_\omega=-\frac{2}{3}\dot\omega/\omega$, $\pi_i=\omega\dot\alpha_i$ conjugate to $\phi,\omega,\alpha_i$.
By Liouville's theorem, the symplectic form $\varpi=d\pi_\phi\wedge d\phi + d\pi_\omega\wedge d\omega + d\pi_i\wedge d\alpha_i$ is time-invariant so its $t\to\pm  \infty$ limits coincide.
The asymptotics $\Lcal\simeq X/2$ and~\eqref{Einstein-sols}, including the Hamiltonian constraint,
give
$
\varpi
\overset{t\to\pm\infty}{=} \pm d\bigl(\bigl(\tfrac{3}{2}|\lambda|^2\bigr)^{1/2} \phi_0^{\pm}/r(\phi_0^{\pm})\bigr) \wedge d\phi_1^{\pm}
+ d\lambda_i \wedge d\nu_i^{\pm}
$,
and these limits must coincide.
At fixed~$\lambda$, this means $\pm d(\phi_0^{\pm}/r^{\pm})\wedge d\phi_1^{\pm}$ are equal, so the map~$\Phi$ is a canonical transformation as stated in~\eqref{Phi-is-canonical} with $\epsilon=\sgn\gamma=-1$.
This establishes the second scattering law~\eqref{law2} for modified-matter bounces.
One can check that $\phi_0^+,\phi_1^+$ only depend on $\phi_0^-,\phi_1^-$, and the scattering map takes the explicit form~\eqref{Sani} given in our model-independent analysis.


\section{Outlook on universality and model-dependence.}

Our notion of singularity scattering map extracts the  macroscopic effects induced by a microscopic model.
Remarkably, based solely on the ultralocality postulate,
we establish a full classification together with universal laws, while leaving room for model-dependence to affect the universe after the bounce.
The universal laws are obeyed by a wide range of models:
\eqref{law1}~continuity of densitized shear~$\Kcirc\sqrt{g}$,
\eqref{law2}~canonical transformation of matter, and \eqref{law3}~directional metric scaling.

In the pre Big Bang scenario our approach \emph{selects} the natural choice of signs $\beta_+=-\beta_-$
and leads us to an anisotropic scattering map characterized by an explicit~$\Phi$; cf.~\eqref{pre-Big-Bang-phi}.
For modified matter models, the map depends on the Lagrangian yet obeys the universal scattering laws (in homogeneous cases at least) and fits in our classification. We shall treat inhomogeneous bounces in~\cite{LLV-3}.

The keys for our classification were the constraint equations and the fact that space derivatives are negligible near the singularity. Our method should generalize to other matter fields, a cosmological constant, bounces that do not asymptote to general relativity, and Penrose's conformal cyclic cosmology~\cite{PenroseCCC1,Tod:2002wd}.
For compressible fluids we find in~\cite{LeFlochLeFloch-3} an interesting interplay between geometric singularities, fluid shock waves, and phase transitions.


\appendix


\begin{thebibliography}{99}
\nonfrenchspacing
\providecommand{\auth}{\textit}

\bibitem{AnderssonRendall}
\auth{L.~Andersson and A.D.~Rendall,}
Quiescent cosmological singularities,
Commun.\ Math.\ Phys.\ 218 (2001), 479--511.

\bibitem{Ashtekar}
\auth{A.~Ashtekar,}
Singularity resolution in loop quantum cosmology: a brief overview,
J.~Phys.\ Conf.\ Ser.\ 189 (2009), 012003. 

\bibitem{AshtekarWilsonEwing}
\auth{A.~Ashtekar and E.~Wilson-Ewing,}
Loop quantum cosmology of Bianchi I models,
Phys.\ Rev.\ D 79 (2009) 083535.

\bibitem{Barrow}
\auth{J.D.~Barrow,}
Quiescent cosmology,
Nature 272 (1978), 211--215.

\bibitem{Bars}
\auth{I. Bars, P.J. Steinhardt, and N. Turok,}
Cyclic cosmology, conformal symmetry and the metastability of the Higgs,
Phys. Lett. B726 (2013), 50--55.

\bibitem{Belbruno:2018uek}
\auth{E.~Belbruno and B.~Xue,}
Regularization of the big bang singularity with random perturbations,
Class.\ Quant.\ Grav.\ 35 (2018) 065013.

\bibitem{BKL} 
\auth{V.A.~Belinsky, I.M.~Khalatnikov, and E.M.~Lifshitz,}
Oscillatory approach to the singular point in relativistic cosmology,
Adv.\ Phys.\ 19 (1970), 525--573.

\bibitem{Berger:2002st}
\auth{B.K.~Berger,}
Numerical approaches to space-time singularities,
Living Rev.\ Rel.\ 5 (2002), 1.

\bibitem{BrandenbergerPeter}
\auth{R.~Brandenberger and P.~Peter,}
Bouncing cosmologies: progress and problems, 
Foundations of Physics 47 (2017), 797--850.

\bibitem{Cai}
\auth{Y.-F.~Cai, A. Marcian\`o, D.-G. Wang, and E. Wilson-Ewing,} 
Bouncing cosmologies with dark matter and dark energy, 
Universe 2017, 3, 1.

\bibitem{Cook:2020oaj}
\auth{W.G.~Cook, I.A.~Glushchenko, A.~Ijjas, F.~Pretorius and P.J.~Steinhardt,}
Supersmoothing through Slow Contraction,
Phys.\ Lett.\ B 808 (2020), 135690.

\bibitem{Damour-et-al}
\auth{T.~Damour, M.~Henneaux, A.D.~Rendall, and M.~Weaver,}
Kasner-like behavior for subcritical Einstein-matter systems,
Ann.\ Henri Poincar\'e 3 (2002), 1049--1111.

\bibitem{Cesare-bounce}
\auth{M.~De Cesare and E.~Wilson-Ewing}
A generalized Kasner transition for bouncing Bianchi I models in modified gravity theories,
J. Cosmo. and Astro. Phys. 1912, no. 12 (2019), 039.

\bibitem{GasperiniVeneziano1}
\auth{M.~Gasperini and G.~Veneziano,}
Pre-Big Bang in string cosmology,
Astro.\ Phys.\ 1 (1993), 317.

\bibitem{GasperiniVeneziano2} 
\auth{M.~Gasperini and G.~Veneziano,}
The pre-Big Bang scenario in string cosmology,
Phys.\ Rep.\ 373 (2003), 1--212.

\bibitem{Ijjas:2020dws}
\auth{A.~Ijjas, W.G.~Cook, F.~Pretorius, P.J.~Steinhardt and E.Y.~Davies,}
Robustness of slow contraction to cosmic initial conditions,
JCAP 08 (2020), 030.

\bibitem{Israel} 
\auth{W.~Israel,} 
Singular hypersurfaces and thin shells in general relativity,
Nuovo Cim.\ 44B (1966), 1--14.

\bibitem{KOST}
\auth{J.~Khoury, B.A.~Ovrut, P.J.~Steinhardt, and N.~Turok,}
The ekpyrotic universe: colliding branes and the origin of the hot Big Bang,
Phys.\ Rev.\ D 64 (2001), 123522.

\bibitem{LeFlochLeFloch-1}
\auth{B.~Le Floch and P.G.~LeFloch,}
On the global evolution of self-gravitating matter. Nonlinear interactions in Gowdy symmetry,
Arch.\ Rational Mech.\ Anal.\  233 (2019), 45--86.

\bibitem{LeFlochLeFloch-2}
\auth{B.~Le Floch and P.G.~LeFloch,}
Compensated compactness and corrector stress tensor for the Einstein equations in $\mathbb{T}^2$ symmetry, 
Preprint ArXiv:1912.12981. 

\bibitem{LeFlochLeFloch-3}
\auth{B.~Le Floch and P.G.~LeFloch,}
On the global evolution of self-gravitating matter.
Phase boundaries, scattering maps, and causality, in preparation.

\bibitem{LeFlochLeFloch-4}
\auth{B.~Le Floch and P.G.~LeFloch,}
On the global evolution of self-gravitating matter, in preparation.

\bibitem{LLV-1}
\auth{B.~Le Floch, P.G.~LeFloch, and G.~Veneziano,}
Cyclic spacetimes through singularity scattering maps, 
Preprint ArXiv:2005.11324. 

\bibitem{LLV-3}
\auth{B.~Le Floch, P.G.~LeFloch, and G.~Veneziano,}
in preparation.

\bibitem{LeFloch-Oslo}
\auth{P.G.~LeFloch,}
Kinetic relations for undercompressive shock waves. Physical, mathematical, and numerical issues, 
Contemp.\ Math.\ 526 (2010), 237--272.


\bibitem{Lehners}
\auth{J.-L. Lehners, P. McFadden, N. Turok, and P.J. Steinhardt,}
Generating ekpyrotic curvature perturbations before the big bang, 
Phys. Rev. D76, 103501 (2007).  

\bibitem{PenroseCCC1}
\auth{R.~Penrose,}
Before the big bang: an outrageous new perspective and its implications for particle physics,
in: ``EPAC 2006 proceedings'',
ed.\ C.R.~Prior, 2006, EPS Accelerator Group, Edinburgh, pp. 2759–2762.

\bibitem{SteinhardtTurok2004}
\auth{P.J.~Steinhardt and N.~Turok,}
Beyond inflation: a cyclic universe scenario,
Phys.\ Scripta T 117 (2005), 76.

\bibitem{Tod:2002wd}
\auth{K.P.~Tod,}
Isotropic cosmological singularities,
in ``The conformal structure of spacetime: Geometry, Analysis, Numerics'', Springer Verlag, 2002, pp.~123--134.

\bibitem{VenezianoSFD}
\auth{G.~Veneziano,}
Scale factor duality for classical and quantum strings,
Phys.\ Letters 265 (1991), 287--294.

\bibitem{Wilson-Ewing-LQC}
\auth{E.~Wilson-Ewing,}
The loop quantum cosmology bounce as a Kasner transition,
Class.\ Quant.\ Grav.\ 35 (2018), 065005.

\bibitem{Xue:2014oea}
\auth{B.~Xue and E.~Belbruno,}
Regularization of the big bang singularity with a time varying equation of state $w > 1$,
Class.\ Quant.\ Grav.\ 31 (2014), 165002.

\end{thebibliography}
\end{document}